\documentstyle[11pt]{article}
\begin{document}
\begin{titlepage}
\vspace{1.5cm}
\title{
{\center \bf Glauber dynamics and ageing}}
\author{
R. M\'elin
\thanks{e--mail: melin@crtbt.polycnrs--gre.fr}
\thanks{Present address: International School for Advanced
Studies (SISSA), Via Beirut 2--4, 34014 Trieste, Italy}
and
P.Butaud
\thanks{e--mail: butaud@crtbt.polycnrs-gre.fr}
{}\\
CRTBT--CNRS, 25 Avenue des Martyrs, BP 166X, 38042
Grenoble CEDEX, France
}
\maketitle
\begin{abstract}
\normalsize
The Glauber dynamics of various models (REM--like trap
models,
Brownian motion, BM model, Ising chain and SK model)
is analyzed in relation with the existence of ageing.
From a finite size Glauber matrix, we calculate
a time $\tau_w(N)$ after which the system has relaxed
to the equilibrium state. The case of metastability
is also discussed.
If the only non zero overlaps between pure states are only
self--overlaps (REM--like trap models,
BM model), the existence or
absence of ageing depends only on the behavior
of the density of eigenvalues for small eigenvalues.
We have carried out a detailed numerical and
analytical analysis of the density of eigenvalues
of the REM--like trap models. In this case, we show
that the behavior of the  density of eigenvalues for
typical trap realizations is related to the spectral
dimension of the equivalent random walk model.
\end{abstract}
\end{titlepage}

\renewcommand{\thepage}{\arabic{page}}
\setcounter{page}{1}
\baselineskip=17pt plus 0.2pt minus 0.1pt
\section{Introduction}
Various approaches have been used to understand ageing
experiments \cite{exp1} \cite{exp2} \cite{exp3}:
droplet picture \cite{droplet}, mean field model \cite{MF1}
\cite{MF2}, and trap models \cite{Bouchaud1} \cite{Bouchaud2}
\cite{Bouchaud3} \cite{Monthus}.
Ageing is by definition the property that the two
points correlation function
\begin{equation}
\label{eqcorr}
C(t_w,t) = \langle {\cal O}(t_w+t) {\cal O}(t_w) \rangle
\end{equation}
depends explicitly on the waiting time $t_w$. ${\cal O}$
is a macroscopic observable, for instance the magnetization
in the zero field cooled experiment \cite{exp1} or the thermoremanent
magnetization experiment \cite{exp2}.
On the theoretical level,
dynamics can be implemented in several ways. One possibility
is Langevin dynamics which was used for instance in \cite{CK} 
where mean field ageing was first obtained, or in \cite{CKP}
where ageing in the presence of flat directions was studied.
Another possibility is Glauber dynamics \cite{Glauber}.
Within the framework of Glauber dynamics and
given the probability distribution $p(\{e_i\},0)$,
for the system to be in any microcanonical state $e_i$ at time $t=0$,
the probability distribution at time $t$ is given by
\begin{equation}
p(\{e_i\},t) = \exp{(T t)} p(\{e_i\},0)
,
\end{equation}
where $T$ is the Glauber matrix.
By "microcanonical states", we mean
a state such as the system is in a given microcanonical
configuration with probability one.

This article is devoted to the study of the Glauber dynamics
of various models,
phenomenological or microscopic, in connection
with ageing.
More precisely, for a finite size system
of size $N$, we calculate from the Glauber matrix
a time $\tau_w(N)$ such as the finite size
system equilibrates on a time of the order
of $\tau_w(N)$, starting from out--of--equilibrium
initial conditions at time $t=0$. In terms of ageing,
the correlation function (\ref{eqcorr}) depends
explicitely on the waiting time $t_w$
provided the system is out--of--equilibrium
at time $t_w$, namely $t_w$ is smaller
than $\tau_w(N)$.
In other words, ageing for a finite size system
is interupted after $\tau_w(N)$ [see \cite{Bouchaud1}
where the denomination ``interrupted ageing''
was proposed]. In order to understand ageing in
the thermodynamic limit ($N \rightarrow \infty$),
the variations of $\tau_w(N)$ with the system size
have to be carefully studied. If $\tau_w(N)$
goes to a constant in the thermodynamic limit,
ageing is interrupted, whereas if $\tau_w(N)$
diverges ageing is not interrupted in the thermodynamic
limit: in the thermodynamic limit,
the system reaches equilibrium on infinite time scales.
We also discuss the possibility of metastability
without ageing on experimental time scales,
as it is the case for instance in diamond.

In general, $\tau_w(N)$ depends on the eigenvalues
of the Glauber matrix, but also on the overlaps between
microcanonical states, and the transformation matrix
elements between the eigenstates of the Glauber
matrix and the microcanonical states.
However, in some simple cases, with only self--overlaping
states, the time $\tau_w(N)$ can be expressed
only in terms of the density of eigenvalues
of the Glauber matrix.

The form of the Glauber matrix will be given
in each case in the core of the paper.
Whatever the system,
the eigenvalues of $T$ are negative. The spectrum
is bounded above by zero (zero is always an eigenvalue,
corresponding to the equilibrium Boltzmann
distribution).
The eigenvalues
of the Glauber matrix have the dimension
of the inverse of a time, so that
they can be viewed as the inverse relaxation
times of the system. Since we are interested in long
time behavior, a special attention will be
paid to the smallest eigenvalues.

This article is organized as follows. In section \ref{GM},
we present a detailed study of the average density of small
eigenvalues of the random energy model (REM) like trap
model proposed and studied in
\cite{Bouchaud1,Bouchaud2,Bouchaud3} and inspired by the
REM model \cite{Derrida}.

In the case of the REM--like trap model,
two regimes are predicted and checked
numerically for the density of small eigenvalues in the glass
phase: an Arrhenius--like regime and a non Arrhenius
regime for the smallest eigenvalues. By Arrhenius--like regime,
we mean a regime in which the density of eigenvalues
is the same as the density of eigenvalues calculated
from the Arrhenius law, assuming a correspondence between
the eigenvalues of the Glauber matrix and the inverse
relaxation times of the system.
The eigenvalue density in the Arrhenius regime
are given in section \ref{Arrh-def}. We also
establish a connection between the behavior of
the density of small eigenvalues of REM--like
trap models and the spectral dimension of the associated
diffusion process.

Next, in section \ref{general}, we give a general relation
for the existence of ageing from a Glauber dynamics
point of view. This section is the main part of
the present article.
When applied to the REM--like trap model,
our criterium is equivalent to the fact noticed in
\cite{Bouchaud1} that the average trapping time diverges
below $T_g$. The rest of the paper is devoted to
examine the case of other models: Brownian motion
which was shown in \cite{CKP} to be an ageing phenomenon,
a model introduced by Barrat and M\'ezard \cite{Mezard}
(BM model), 
the one dimensional Ising chain (in this case, our results are
consistent with the existence of interrupted ageing),
and the Sherrington--Kirkpatrick model (SK model).
In all these cases, we find consistent results between the existence
of ageing calculated from the behavior of the two times autocorrelation
functions and our criterium based on the Glauber matrix.
We end--up with some final remarks.
\section{Density of small eigenvalues of trap models
(REM--like trap models and trap models with other topologies)}
\label{GM}
Trap models can be defined without a priori reference
to a microscopic model. However, some trap models,
such as the REM--like trap model \cite{Bouchaud3}
are related
to the Parisi solution of spin glasses: even though
these models are toy models, some of them retain some
feature of the dynamics of microscopic models.
The REM--like trap model consists of a trap model in infinite
dimension, that is, the system can "hop" from one
trap to any other trap, according to the rules of the dynamics.
Other toplogies will be analyzed below, in reduced dimension.
The REM--like trap model corresponds to a single
replica symmetry breaking step. Generalizations
of the REM--like trap model including multi
replica symmetry breaking
steps were studied in \cite{Bouchaud3} and their relevance
to explain experimental data was shown.

The models of this section consist of $N$ traps
of depth $E_i<0$, with the $E_i$ chosen among the
distribution $P(E)=\exp(E)$. The reason why
this particular energy trap distribution is chosen
is that
there exists a finite transition temperature between
a high and a low temperature glassy phase \cite{Monthus}.
The dynamics rules are given by
$T_{i,j} \propto \exp{(\beta E_j)}$ for $i \ne j$
if the traps $i$ and $j$ are connected and zero otherwise.
The diagonal terms are defined by
\begin{equation}
\sum_{k=1}^{N} T_{k,i}=0
\label{propcons}
,
\end{equation}
which enforces probability conservation.
Equation (\ref{propcons}) is a general constraint for
any Glauber dynamics. Another constraint
is the detailed balance condition
\begin{equation}
T_{j,i} \exp{(-\beta E_i)} = T_{i,j} \exp{(- \beta E_j)}
.
\end{equation}
It is clear that the aforementioned rules for the dynamics
of the REM--like trap model satisfy detailed balance.
However, as we
shall see below, other types of dynamics also satisfy detailed
balance and probability conservation.
As a consequence of detailed balance,
the eigenvectors of the Glauber matrix are not physical
states (in the sense that they are not probability distributions)
except for the Boltzmann distribution (the stationary eigenvector
corresponding to the $\lambda=0$ eigenvalue).
Notice that if $T$ is a Glauber matrix $x T$, with
$x$ a positive number, is also a Glauber matrix.
Multiplying $T$ by $x$ amounts to changing the unit time.
The proportionality constant is imposed by physical
considerations. For the models considered
in this section, this constant
is chosen such as the diagonal elements of the Glauber
matrix do not scale with the number $N$ of traps in the large $N$
limit. For instance, in the infinite dimensional case, we choose
$T_{i,j} = \exp{(\beta E_j)}/N$.
This dynamics will be studied with different phase space topologies:
infinite dimension, where all the traps are connected (section
\ref{infinite}), one dimensional
topology, where a given trap is connected only to its two
nearest neighbors (section \ref{disor}),
and also one dimensional topology with ``ordered'' traps,
in the sense that the trap energies have been ordered
$E_1<...<E_N$, and the trap at site $i$ has an energy $E_i$.

A useful property that we will use throughout this paper
is that Glauber matrices can
be mapped onto symmetric matrices with the same spectrum.
Let us call $V$ the diagonal matrix
$V_{i,j} = \exp{(\beta E_i/2)} \delta_{i,j}$.
Then, because of detailed balance, the matrix
$\tilde{T} = V T V^{-1}$ is symmetric and has the same spectrum
as $T$.
Using a quantum mechanical analogy, we denote by $|e_i \rangle$
the microcanonical
state corresponding the system in trap $i$ with
a probability unity.
The analogy to quantum mechanics is useful, even though
there is no hermitian structure. In particular, any linear combination
of microcanonical states is not
a probability distribution.
\subsection{Arrhenius regime}
\label{Arrh-def}
We propose ourselves to carry out a detailed study of the
density of eigenvalues of the
REM--like trap model and related models
in reduced dimension. Since the Arrhenius law
is introduced "by hand" in the dynamics,
it is interesting to question in which
regime the density of eigenvalues is given by the
Arrhenius law, assuming a correspondence $\tau=-1/\lambda$
between the eigenvalues of the Glauber matrix and relaxation times.
We will naturally call {\it Arrhenius regime} such a regime.
As we will see below, the density of eigenvalues is not always in
the Arrhenius regime.

In this Arrhenius regime, and by definition,
the off-diagonal terms are neglectable as far
as the density of eigenvalues is concerned.
The diagonal coefficients of the Glauber matrix are
proportional to $\exp(\beta E_i)$, so that the distribution
of eigenvalues is
\begin{equation}
P(\ln{|T|}) \propto \exp{(\frac{1}{\beta} \ln{|T|})}
\label{Aha}
.
\end{equation}
This distribution is the same as the one expected from
an Arrhenius law with life times
$\tau_i \propto \exp{(- \beta E_i)}$. If one assumes an
Arrhenius picture, the distribution of relaxation times
is
\begin{equation}
P(\tau) \propto \frac{1}{\tau^{1+1/\beta}}
,
\end{equation}
which is equivalent to (\ref{Aha}).
As noticed in \cite{Bouchaud3},
the average trapping
time diverges if $\beta>1$, leading to
ageing. Distribution of trapping times with no
first moment were already studied in \cite{Machta},
and shown to lead to anomalous diffusion,
namely to a non trivial spectral dimension,
even above the upper critical dimension.
We will compare our results for the
REM--like trap model with the results
of \cite{Machta} derived of the context
of anomalous diffusion on site--disordered
hypercubic lattices.

The density of eigenvalues is thus
described by the Arrhenius law
provided the off-diagonal terms of the Glauber matrix
are a small perturbation, so that the density of
eigenvalues of the Glauber matrix is given by the diagonal
terms only. We now examine several cases with different
phase space topologies and determine in the different cases
on which condition the density of eigenvalues is determined
by the Arrhenius picture. We also compare our analytical
approach to numerical diagonalizations of the symmetrized matrix
$\tilde{T}$.
\subsection{Infinite dimensional case: REM--like trap model}
\label{infinite}
In this section, the system can hop from one trap
to any other trap (infinite dimension).
The diagonal coefficients are
\begin{equation}
T_{i,i}=\tilde{T}_{i,i}=-\frac{N-1}{N} \exp{(\beta E_i)}
,
\end{equation}
and the off--diagonal coefficients of the symmetrized Glauber matrix
are
\begin{equation}
\tilde{T}_{i,j} = \frac{1}{N} \exp{(\frac{\beta}{2}(E_i+E_j))}
.
\end{equation}
\subsubsection{Determination of the Arrhenius regime}
We treat the off--diagonal coefficients $\tilde{T}_{i,j}$
($i \ne j$) as a perturbation.
This perturbative treatment
is consistent since we are looking for a regime in which
the off--diagonal coefficients do not modify the density of
eigenvalues with respect to the purely diagonal case.
The second order correction to the eigenvalue $T_{i,i}$ is
\begin{equation}
\delta^{(2)}_i = \frac{1}{2 N (N-1)}
\sum_{k \ne i} \frac{\exp{(\beta(E_i+E_k)/2)}}
{\sinh{(\beta(E_k-E_i)/2)}}
,
\end{equation}
the first order corrections being identically zero since
the perturbation consists only of off--diagonal elements.
We assume that the density of trap energies in the
vicinity of $E_i$ is large enough for the discrete summation
to be replaced by an integral. If this condition is
not satisfied, the second order corrections $\delta^{(2)}_i$
may have large fluctuations
because the divergence for $E_k \rightarrow E_i$
is no more cancelled. The condition that the unperturbed
eigenvalue density is large in the vicinity of $E_i$ simply
reads $N \exp{(E_i)} \gg 1$. We assume in the rest of this section
that this condition is fulfilled, and we can safely replace
the discrete sum by an integral:
\begin{equation}
\delta^{(2)}_i = \frac{1}{2(N-1)} \int_{- \infty}^{0}
\frac{\exp{(\beta(E_i+E)/2)}}
{\sinh{(\beta(E-E_i)/2)}} e^{E} dE
.
\end{equation}
The singularities for $E=E_i^{+}$ and $E=E_i^{-}$ cancel each
other, and we can approximate $\delta^{(2)}_i$ as
\begin{equation}
\label{deltai}
\delta_2^{(i)} \simeq \frac{1}{N-1} B \exp{\left((1+\beta)E_i
\right)}
,
\label{app}
\end{equation}
with $B=4(1+\beta/2)/\beta^{2}$.
We have used the following approximation: if $A$ is small, then
\begin{equation}
\int_{-A}^{A} \frac{
\exp{(\lambda x)}}{\sinh{(\mu x)}} \simeq \frac{2 \lambda A}{\mu}
,
\end{equation}
and we use $A=2/\beta$ as a cut--off. Clearly (\ref{app})
is only an approximation, but we are only interested in the
dominant behavior of $\delta_i^{(2)}$ as a function
of $E_i$.
Now, we want to compare the approximation (\ref{app})
of $\delta^{(2)}_i$ to the typical level spacing between
$T_{i,i}$ and $T_{i+1,i+1}$, where we assume that the
traps are ordered such as $E_1<E_2<...<E_N$.
To to so, we first calculate the distribution of
\begin{equation}
|T|=\frac{N-1}{N} \exp(\beta E)
,
\end{equation}
and we obtain $P(|T|)=A |T|^{1/\beta-1}$, with
\begin{equation}
A = \frac{1}{\beta} \left(
\frac{N}{N-1} \right)^{1/\beta}
.
\end{equation}
Notice here that $P(|T|)$ diverges as $|T|$ goes to
zero if $\beta>1$. This accumulation of small
relaxation times is related to the existence
a glass transition for this model for $\beta=1$.
Let us now calculate the level spacing statistics
$P_{|T|}(S)$, which is the density probability that
the spacing between the absolute value of the eigenvalue $T$
and the absolute value of the next eigenvalue is $S$.
Because of the statistical independence of the trap depths,
$P_{|T|}(S)$ is readily obtained as
\begin{equation}
P_{|T|}(S) \delta S = \sum_{k=1}^{N-1} {N-1 \choose k}
\left(\delta S P(|T|+S) \right)^{k}
\left( \int_0^{|T|} + \int_{|T|+S+\delta S}^{1}
P(t) dt \right)^{N-k-1}
.
\end{equation}
In this expression, $k$ is the number of levels in
the interval $[|T|+S,|T|+S+\delta S]$ (with $k \ge 1$).
The other terms correspond
to the condition that there is no eigenvalue
in the interval $[|T|,|T|+S]$.
Assuming that $|T|$ is small compared to unity, and that
$N$ is large, we get
\begin{equation}
P_{|T|}(S) \simeq \frac{1}{\beta} N (|T|+S)^{1/\beta-1}
.
\end{equation}
The off-diagonal terms are irrelevant to the density
of eigenvalues provided
\begin{equation}
{\cal P}(|T|) = \int_0^{\delta^{(2)}_i}
P_{|T|}(S) dS
\end{equation}
is much smaller than unity. This condition reads
$B e^{2 E_i}/\beta \ll 1$, with $B$ defined in
(\ref{deltai}). Since we are looking for a
necessary condition valid if $\beta>1$, we neglect the
temperature--dependent prefactor and
conclude that the density of states
is in the Arrhenius regime only for traps such that
$1/N \ll e^{E} \ll 1$. Notice that the traps such as
$N \exp{E} \ll 1$ are deeper than the typical deepest trap
(of the order of $-\ln{N}$). The Arrhenius picture
is thus valid for a {\it typical} trap realization, with
a lowest trap not deeper than $-\ln{N}$. However, non
typical disorder realizations may be generated. For these
realizations, the lowest eigenvalues are not described
by an Arrhenius picture. The existence or absence
of the Arrhenius regime even for a very large number of
traps is thus sample dependent. This non self averaging
behavior is reminiscent on the sample dependence of
the quantity
\begin{equation}
Y = \sum_{i=1}^{N} W_i^{2}
,
\end{equation}
with $W_i$ the probability to find the system in trap $i$
\cite{DT}.
\subsubsection{Numerical tests}
The density of small eigenvalues of the REM--like trap
model is pictured on figure \ref{Fig1}.
For practical purposes, we do not
calculate the density $P(\ln{|\lambda|})$ since this quantity
is too small for small eigenvalues (see (\ref{Aha})).
We rather calculate
$P_B(\ln{|\lambda|})= W_B P(\ln{|\lambda|})$, where
$W_B$ is the Boltzmann weight associated to the
eigenvalue $\lambda$, which is calculated as the average
of the Boltzmann weight operator
\begin{equation}
\hat{W}_B = \sum_{i=1}^{N} |e_i \rangle e^{-\beta E_i}
\langle e_i |
\end{equation}
over the eigenstate
associated to the eigenvalue $\lambda$. In the Arrhenius
picture, we have
\begin{equation}
\label{PB}
P_B(\ln{|\lambda|}) \propto \exp{\left(
\left(\frac{1}{\beta}
-1 \right) \ln{|\lambda|}\right)}
.
\end{equation}
Since $\beta>1$, $P_B(\ln{|\lambda|})$ increases exponentially
as a function of $\ln{|\lambda|}$ when
$\lambda \rightarrow 0^{-}$.
We see on figure \ref{Fig1} the existence of the Arrhenius
regime, in good agreement with the condition
$1/N \ll e^{E} \ll 1$. We can now fit the density
of eigenvalues in the Arrhenius regime to the form
\begin{equation}
P_B(\ln{|\lambda|}) \propto \exp{(- \alpha \ln{|\lambda|})}
\label{lawArr}
,
\end{equation}
and see if we recover $\alpha=1-T$, as predicted from the
Arrhenius law. The fit to the Arrhenius law is shown on figure
\ref{Fig2}.
The variations of the exponent $\alpha$ as a function
of the inverse temperature are plotted on figure \ref{Fig3},
and we conclude to a good agreement with the Arrhenius law.
\subsubsection{Relation with the anomalous diffusion
on site--disordered lattices}
\label{Relation1}
Random walks on a finite $d$--dimensional lattice
with a broad distribution of waiting
times have been analyzed in the past
(see \cite{Machta} and references therein).
As we shall see, we can relate our results
concerning the existence of a Arrhenius--like
regime for the REM--like trap model to the results
of \cite{Machta}, where the long time behavior
of random walk on such lattices was investigated
by means of the real space renormalization method
introduced in \cite{Machta2}.
The REM--like trap model falls
in the universality class described by Machta
since the ingredients are the same in the two cases,
rather than the universality class of random
walks in a random field force \cite{Random-field-force}.
In order to make contact with the work
of Machta, we first calculate the probability
$P_0^{(REM)}(t)$ of being at the origin:
\begin{equation}
P_0^{(REM)}(t) = \overline{\frac{1}{N} \sum_{i=1}^{N}
{\cal P}(e_i,t|e_i,0)} =
\overline{\frac{1}{N} \sum_{\alpha} e^{\lambda_{\alpha}t}}
,
\end{equation}
where ${\cal P}(e_i,t|e_i,0)$ is the probability
to find the system in the trap $i$ at time $t$,
starting from the trap $i$ at time $0$, and
$\{\lambda_{\alpha}\}$ is the set of eigenvalues
of the Glauber matrix. We have shown that,
in the infinite dimensional case, the typical
small eigenvalue density (relevant to the
long time behavior of $P_0(t)$) has
the Arrhenius form $\rho(\lambda) \sim
\lambda^{T-1}$. A simple scaling analyzis
then shows that
\begin{equation}
\label{P0REM}
P_0^{(REM)}(t) \sim t^{-T}
\end{equation}
in the long time limit.
This result was derived for a graph such as
any site is connected with any other site,
that is we started from the begining from
an infinite dimensional situation.
In order to extrapolate this result to
a finite dimension $d$ larger than the
upper critical dimension 2 (as shown in \cite{Machta})
we first notice that, within the framework of
the REM--like trap model, the microscopic time
$\tau_0$ scales like $\tau_0^{(REM)} \sim a^{d}$,
with $a$ the lattice spacing, whereas, in the
context of random walks, the correct scaling
is $\tau_0^{(RW)} \sim a^{2}$ [an illustration
of the validity of such a scaling will be given
in section \ref{Brownian-motion}]. This implies
that $\tau_0^{(RW)} \sim (\tau_0^{(REM)})^{2/d}$,
which gives the correct prescription to
go from the REM--like trap model in infinite
dimension to the finite $d$ dimensional
random walk model: one should replace
the time $t$ in (\ref{P0REM}) by $t^{d/2}$.
This leads to
\begin{equation}
P_0^{(RW)} \sim t^{-T d/2}
\end{equation}
for $d$ larger than the upper critical dimension
$2$. This is exactly what was found in
\cite{Machta} using a real space renormalization
group approach, with a non trivial spectral
dimension $d_s=T d$.

\subsubsection{Structure of the eigenvectors of the REM--like
trap model}
\label{eigI}
We aim to calculate the structure of the eigenstates
of the REM--like trap model. To do so,
we decompose a given microcanonical
state $|e_i \rangle$ into the
eigenvectors $|\psi_j \rangle$,
and calculate the absolute value of the coefficients of this
decomposition. The result is plotted on figure \ref{Fig4}.
We see that a microcanonical state
$|e_i \rangle$ with an eigenvalue
$\lambda_i$ is mainly a linear
combination of eigenstates with eigenvalues close
to $\lambda_i$. This observation can be understood
in the framework of perturbation theory as follows.
The first order correction to the eigenvectors are
\begin{equation}
|\delta \phi_i \rangle =
\frac{1}{2(N-1)} \sum_{k \ne i} \frac{1}
{\sinh{(\beta(E_k-E_i)/2)}} |e_k \rangle
,
\label{per1}
\end{equation}
so that, at first order
\begin{equation}
|e_i \rangle = |\psi_i \rangle - \frac{1}{2(N-1)}
\sum_{k \ne i} \frac{1}{\sinh{(\beta(E_k-E_i)/2)}}
|\psi_k \rangle
.
\label{per2}
\end{equation}
At the first order of perturbation theory, the
pure states $|e_i \rangle$ is thus mainly a linear
combination of eigenvectors with an eigenvalue
close to the eigenvalue of $|\psi_i \rangle$.

Starting from a pure state $|e_i \rangle$ at time
$t=0$, we can
calculate the evolution of the system:
\begin{eqnarray}
|e_i(t) \rangle &=& \exp{\left(-\frac{N-1}{N} e^{\beta E_i} t
\right)} |e_i \rangle \\
&& + \frac{1}{2(N-1)}
\sum_{k \ne i} \frac{1}{\sinh{(\beta(E_k-E_i)/2)}}
\left[ \exp{\left(-\frac{N-1}{N}e^{\beta E_i } \right)}
- \exp{\left(- \frac{N-1}{N} e^{\beta E_k} t \right)} \right]
\nonumber
,
\end{eqnarray}
where we have used (\ref{per1}) and (\ref{per2}).
As expected from the Arrhenius law, the relaxation time
is of the order of $\tau_i = \exp{(-\beta E_i)}$,
which diverges in the limit $\beta \rightarrow + \infty$.
We will compare below the structure of the eigenvectors
of the REM--like trap model to the structure of eigenvectors
of another model (BM model).

\subsection{One dimensional case with disordered traps}
\label{disor}
\subsubsection{Absence of Arrhrenius regime}
Assuming periodic boundary conditions,
the diagonal terms of the Glauber matrix are given by
\begin{equation}
T_{i,i} = -2 \exp{(\beta E_i)}
.
\end{equation}
The perturbation induced by the off-diagonal terms does not
affect significatively the density of eigenvalues provided
the perturbation $\tilde{T}_{i,j}=\exp{(\beta(E_i
+ E_j)/2)}$ is small compared to the level spacing between
$T_{i,i}$ and $T_{i+1,i+1}$.
Since $E_j$ is uncorrelated from $E_i$, the matrix element
$T_{i,j}$ must be averaged over $E_j$, so that
$\tilde{T}_{i,j}$ is of the order of $\exp{(\beta E_i/2)}$,
that is
\begin{equation}
\tilde{T}_{i,j} \sim \sqrt{\frac{|T_{i,i}|}{2}}
.
\end{equation}
The distribution of eigenvalues
of the unperturbed system is
\begin{equation}
P(|T|) = \frac{1}{2^{1+1/\beta}} |T|^{1/\beta-1}
,
\end{equation}
from what we deduce the level spacing statistics in the
long time (small $|T|$) approximation
\begin{equation}
P_{|T|}(S) \simeq \frac{N}{\beta} (|T|+S)^{1/\beta-1}
.
\label{P(S)}
\end{equation}
The off-diagonal terms do not affect the density of eigenvalues
provided
\begin{equation}
\int_0^{\sqrt{|T|/2}} P(S) dS \ll 1
,
\end{equation}
which leads to the condition $N^{2} \exp{E} \ll 1$.
The average number of traps generated among a sequence
of $N$ independent traps chosen among the distribution
$P(E)=\exp{E}$ and fulfilling the aforementioned condition is
\begin{equation}
N \int_{- \infty}^{-2 \ln{N}} e^{E} dE = \frac{1}{N}
,
\end{equation}
which is vanishing in the large $N$ limit. We thus conclude
that the Arrhenius regime with this topology does not exist,
that is off-diagonal terms will always modify significatively
the density of eigenvalues with respect to the Arrhenius case.
As we will see below, the Arrhenius regime is replaced by
a renormalized Arrhenius regime, where the density of eigenvalues
is given by (\ref{lawArr}), but with $\alpha$ lower than $1-T$.
\subsubsection{Numerical tests}
Again, we generate numerically the density of small eigenvalues
$P_B(\ln{|\lambda|})$. The fit is shown on figure \ref{Fig2},
and the variations of the exponent in the Arrhenius regime
are plotted on figure \ref{Fig3}. We conclude to the
existence of a renormalized Arrhenius regime, in the
sense that the density of eigenvalues is given by
(\ref{lawArr}) but with an exponent $\alpha$
smaller than $1-T$, indicating a slowing down of the dynamics
due to the one dimensional topology compared to
the infinite dimensional case: in one dimension,
to go from one trap to
a deeper trap, the system has to jump to intermediate traps.
\subsubsection{Relation with the anomalous diffusion
on site--disordered lattices}
Section \ref{Relation1} was devoted to the connection
between the spectrum of the Glauber
matrix in infinite dimension and the long
time behavior of random walks on finite
$d$ dimensional trapped hypercubic lattices
above the upper critical dimension. We now analyze
the analogy in one dimension. We have shown numerically
that the density of eigenvalues behaves like
$\rho(\lambda) \sim \lambda^{1-T'}$ for small $\lambda$,
where $T'$ is different from
$T$. In terms of random walks, this result
becomes, with the notations of section \ref{Relation1}
\begin{equation}
P_0^{(RW)}(t) \sim t^{-T'/2}
,
\end{equation}
namely the spectral dimension is $d_s=T'$.
We now use the results of \cite{Machta} where
$d_s$ was calculated:
\begin{equation}
T' = d_s = 2 \nu = \frac{2 T}{1+T}
,
\end{equation}
from what we conclude that the coefficient
$\alpha$ plotted on figure \ref{Fig3} is
\begin{equation}
\alpha=\frac{1-T}{1+T}
.
\end{equation}
This analytic form is  
consistent with numerical calculations
of the density of eigenvalues.

\subsection{One dimensional case with ordered traps}
We now assume that $N$ energy traps were generated among
the distribution $P(E)=\exp{E}$, and that these energies
have been ordered in such a way that $E_1<E_2<...<E_N$.
A site $k$ with a trap of energy $E_k$ is connected to
the sites $k-1$ (with an energy $E_{k-1}$) and $k+1$ (with
an energy $E_{k+1}$, except for the sites $1$ and $N$
which have no neighbor. Intuitively, such a trap
ordering should make the dynamics faster
with respect to the case analyzed in section \ref{disor}
since the system can decrease its energy continuously during
the relaxation. We have
\begin{equation}
\tilde{T}_{i,i+1} = \frac{1}{2} \sqrt{T_{i,i} T_{i+1,i+1}}
.
\end{equation}
Since the density of level spacings $P_{|T|}(S)$
(\ref{P(S)})
is large for small values of $|T|$, we expect that
$T_{i+1,i+1}$ is close to $T_{i,i}$ at least if
$|T_{i,i}|$ is small, so that we can use the following
approximation: $\tilde{T}_{i,i} = T_{i,i}/2$.
The off-diagonal terms will not modify the density
of eigenvalues with respect to the Arrhenius case provided
\begin{equation}
\int_0^{|T|/2} P_{|T|}(S) dS \ll 1
,
\end{equation}
that is $N \exp{E} \ll 1$.
The number of traps fulfilling this condition
in a distribution of $N$ traps is on average
equal to unity, so that
only a small number of eigenvalues contribute to the
Arrhenius regime. Nevertheless, it is still possible
to make a statistics over a large ensemble of $N$ traps
distribution
and to test numerically if the Arrhenius regime can be observed.
\subsubsection{Numerical tests}
We generated numerically the density of small eigenvalues
$P_B(\ln{|\lambda|})$.
As shown on figure \ref{Fig2},
we still can fit $P_B(\lambda)$ to the form $\exp{(
- \alpha \ln{|\lambda|})}$, and the exponent $\alpha$
is close to $1-T$, as in the Arrhenius regime.
(see figure \ref{Fig3}).
We conclude that ordering the traps
in one dimension restores the Arrhenius
regime for the lowest eigenvalues. Again,
the existence of the Arrhenius regime is sample dependent.
\section{General relation between ageing and the Glauber matrix}
\label{general}
We are now going to give a general relation between
ageing and the Glauber matrix. This relation is a generalization
of a criterium given in \cite{Bouchaud3}. In the context of
the REM--like trap model, the average trapping time
\begin{equation}
\langle \tau \rangle = \tau_0 \int_{- \infty}^{0}
dE \exp{(E(1-\beta))}
\end{equation}
diverges if $T<T_g=1$. On the other hand, ageing exists
below $T_g$ only, in the limit where the microscopic
time scale $\tau_0$ goes to zero. We are now going to
argue that a similar criterium can be found for
any statistical mechanical model, provided the
master equation is linear. This is not always the case.
For instance, the master equation of the Backgammon model
in a mean field dynamics
at zero temperature is not linear \cite{backgammon}.

Let us now start our general argument. The idea is
to start with a system in a given microcanonical
state $|e_i \rangle$
and write the Glauber dynamics equation for the evolution
of this pure state. First, we call $P$ the transformation
matrix
\begin{equation}
|e_i \rangle = \sum_{\alpha} P_{i,\alpha} | \Psi_{\alpha}
\rangle
\label{pure2}
,
\end{equation}
where $|e_i \rangle$ is a microcanonical state and
$\{|\Psi_{\alpha} \rangle\}$
are the eigenstates of the Glauber matrix.
By microcanonical state, we mean that if the
system is in the state $|e_i \rangle$,
the system is with probability one
in the microcanonical state $e_i$.
In order to avoid confusion, all the labels related to
the eigenvectors will be denoted by greek letters whereas the
labels for pure states will be denoted by latin letters.
The evolution of the pure state (\ref{pure2}) is readily
obtained as
\begin{equation}
\label{state}
|e_i(t) \rangle = \sum_{\alpha,j} P_{i,\alpha}
e^{\lambda_{\alpha}t} P_{\alpha,j}^{-1}
.
\end{equation}
In other words,
the probability to be in microcanonical state
$e_j$ at time $t$
starting
from the microcanonical state $e_i$ is
\begin{equation}
{\cal P}(e_j,t|e_i 0) = \sum_{\alpha}
P_{i,\alpha} e^{\lambda_{\alpha} t} P_{\alpha,j}^{-1}
.
\end{equation}
We now consider the time--dependent correlation function
\begin{equation}
{\cal C}_N(t) = \frac{1}{{\cal N}} \sum_{i,j}
{\cal P}(e_j,t|e_i,0) q_{i,j}
\label{correlation}
,
\end{equation}
where ${\cal N}$ is the number of
microcanonical states. We use
${\cal N}$ instead of $N$ to avoid confusion between the
number of microstates ${\cal N}$ and the number of sites $N$
in a spin model. In a trap model, ${\cal N}=N$,
whereas in an Ising spin system, ${\cal N}=2^{N}$.
In (\ref{correlation}), we have carried out an
average over all the initial microcanonical
states, that is
the system is first quenched from an infinite temperature,
and the correlation (\ref{correlation}) is averaged over
the realizations of the infinite temperature states.
In (\ref{correlation}), $q_{i,j}$ is the overlap
between the pure states $e_i$ and $e_j$,
which has to be defined within each model.
As we will see below in section \ref{Brownian-motion},
even within a given dynamics, the choice of the
overlap depends on what type of observables
are considered in the two times correlation functions.
We now want to calculate from (\ref{correlation}) a
typical time by integrating (\ref{correlation}).
However, the Boltzmann
eigenstate leads to a trivial divergence
of this integral.
We thus need a prescription to eliminate
the equilibrium distribution in (\ref{correlation}).
We first discuss this prescription in the absence
of metastability and postpone for later
discussion the case of metastability.
We define the typical time $\tau_w$ by
\begin{equation}
\label{prescription}
\tau_w(N) = \frac{1}{{\cal N}}
\overline{\sum_{i,j} \sum_{\alpha \ne \mbox{Bol.}}
P_{i,\alpha} \frac{1}{\lambda_{\alpha}}
P_{\alpha,j}^{-1} q_{i,j} }
,
\end{equation}
where the summation over the eigenstates excludes the
Boltzmann distribution. The overline denotes an average
over the disorder realizations, if necessary.
The significance of $\tau_w(N)$
is clear: for a finite size system with $N$ sites,
the out--of--equilibrium dynamics
occurs before $\tau_w(N)$ and, for times
larger than $\tau_w(N)$, the finite
size system has equilibriated. In other words,
$\tau_w(N)$ is the cross--over time before
which the finite size system exhibits
ageing.
Notice that in this argument, the system is a finite
size system. One should thus carefully study finite size
effects. Two cases may occur: first, $\tau_w(N)$ tends
to a constant in the thermodynamic limit. In this case,
ageing is interrupted \cite{Bouchaud1} in the thermodynamic
limit. As we
shall see, this is the case for the
Ising chain at a finite temperature.
On the
other hand, if $\tau_w(N)$
diverges in the thermodynamic limit,
ageing is not interrupted.
In this case, the exitence of a divergent
relaxation time with respect to times $0$ and $t$
implies the existence of a divergent relaxation time
with respect to any finite waiting time $t_w$
and $t+t_w$, and thus the existence of ageing
in the thermodynamic limit. The relation between
the existence/absence, the nature
of ageing in the thermodynamic
limit and the behavior of $\tau_w(N)$ was derived
on heuristic basis. We could not find a
rigourous derivation using a general argument.
However, in what follows, we investigate
the behavior of several models (REM--like
trap model, random walks, BM model, one dimensional
Ising chain and SK model) and we find in all these
cases that our criterium based on the behavior of $\tau_w(N)$
is correct.
As far as non interrupted ageing in the thermodynamic limit
is concerned, we will see below that this is
the case for the BM and SK models. This is also the case
for the REM--like trap model. We have shown in section \ref{infinite}
that for all the typical realizations of the trap
configurations (typical is defined in section \ref{infinite}),
the density of eigenvalues is given by the Arrhenius law,
which means that, in the case of the REM--like trap model, our criterium
for the existence of ageing is equivalent to the criterium
of Bouchaud \cite{Bouchaud1} that we have recalled
at the beginning of this section, provided the overlap
between traps is taken to be $q_{i,j} = q_{EA} \delta_{i,j}$,
where $q_{EA}$ is the Edwards--Anderson order parameter
(within the one replica symmetry breaking step picture).
Notice that the existence of a non Arrhenius regime
generated by non typical trap configurations does not
modify our conclusion regarding the divergence
of $\tau_w^{REM}(\infty)$ since the non Arrhenius regime
only occurs for traps such as $E \le - \ln{N} \rightarrow
- \infty$ if $N \rightarrow \infty$.
The correlation ${\cal C}_{\infty}^{REM}(t)$
is
\begin{equation}
{\cal C}_{\infty}^{REM}(t) = \int_{- \infty}^{0} dE
e^{E} \exp{\left(-t e^{\beta E} \right)}
\label{corrREM}
,
\end{equation}
which is non integrable if $\beta \le 1$.
The typical time
\begin{equation}
\label{tauREM}
\tau_w^{REM}(N) = -\frac{q_{EA}}{N}
\overline{\sum_{\alpha \ne \mbox{Bol.}}
\frac{1}{\lambda_{\alpha}}} = -q_{EA}
\int_{-\infty}^{0}\frac{1}{\lambda}
\rho_N(\lambda) d \lambda
\end{equation}
only depends on the density of
eigenvalues $\rho_N(\lambda)$ for a system of $N$ traps,
and is independent on the transition matrix,
which motivates the study of the density of eigenvalues
carried out in section \ref{GM}.
In general, in any model with only self--overlapping
pure states, the existence of ageing is coded only in the
behavior of the density
of small eigenvalues: there is ageing in the thermodynamic
limit if $\rho_{\infty}(\lambda)$ tends to
a constant if $\lambda$ goes to zero (an exemple
will be provided in section \ref{BM-model})
or if $\rho_{\infty}(\lambda)$ diverges if
$\lambda$ goes to zero (this is
the case for the REM--like trap model
already studied in section \ref{GM} ).

Let us now discuss the behavior of the correlation
function (\ref{correlation}) in the presence
of metastability. We have in mind the case
of diamond for instance, which is metastable
but in which case there is no ageing. We do not
have a Glauber matrix description of diamond,
but take this case as a generic situation
for metastability and draw qualitative conclusions.
In this
case we expect that for a finite size system,
(\ref{correlation}) will first decay to
a finite value over a first time scale
$\tau_w^{(1)}(N)$. After $\tau_w^{(1)}(N)$,
the system reaches its metastable configuration
(for instance diamond) and thermalizes in a
portion of its phase space. We thus expect
that, for times larger than $\tau_w^{(1)}(N)$,
the correlation (\ref{correlation}) reaches
a plateau. On larger time scales,
the correlation (\ref{correlation}) should
decay to its equilibrium value with a cross--over
time $\tau_w^{(2)}(N)$, corresponding
for instance to the transition from diamond to
graphite. The time scale $\tau_w^{(2)}(N)$
should increase with the number of sites,
so that, with a macroscopic number of sites,
no transition from diamond to graphite
is observable on experimental time scales.
On the other hand, $\tau_w^{(1)}(N)$
may remain small even in the thermodynamic
limit, signaling the absence of
ageing in the graphite
phase. In such a situation where the cross--over
time scale $\tau_w^{(2)}(N)$ is not
observable, the only relevant time scale
is $\tau_w^{(1)}(N)$. In order to calculate
$\tau_w^{(1)}(N)$ one should not use the
prescription (\ref{prescription}) but rather
substract to the correlation (\ref{correlation})
its plateau value for $\tau_w^{(1)}(N) \ll t
\ll \tau_2^{(2)}(N)$. 
\section{Brownian motion}
\label{Brownian-motion}
Brownian motion was shown in \cite{CKP} to be an
ageing phenomenon. This model is simple and thus provides
a simple test of the criterium of section \ref{general}. Brownian
motion in an euclidian space of dimension $d$ has the
following Glauber dynamics: $T_{i,i}=-1/\tau_0$ and $T_{i,j}
= 1/(2 d \tau_0)$ if $i$ and $j$ are nearest neighbors
on a hypercubic lattice.
$\tau_0$ is the microscopic hopping time.
In this case, the Glauber matrix is symmetric and, because
of translational invariance, its eigenstates are plane
waves
\begin{equation}
| \Psi_{\bf k} \rangle =
\frac{1}{\sqrt{N}} \sum_{\bf x} e^{i {\bf k}.{\bf x}}
| \Psi_{\bf x} \rangle
,
\end{equation}
where ${\bf k} = (n_1,...,n_d) 2 \pi/L$
(we have assumed
cyclic boundary conditions with a lattice spacing $a$).
The matrix elements of the Glauber matrix on the plane
wave basis are
\begin{equation}
\langle \Psi_{\bf k}|T|\Psi_{\bf l} \rangle
=  \delta_{{\bf k},{\bf l}}
\frac{1}{\tau_0}
\left( -1 + \frac{1}{d} \sum_{\alpha=1}^{d}
\cos{(k_{\alpha} a)} \right)
,
\end{equation}
where $\alpha$ runs over the $d$ directions and $k_{\alpha}$
is the component of the wave vector ${\bf k}$ on the $\alpha$
direction. The long time behavior of the random walk is readily
obtained by expanding the cosine up to the second order, and
the probability to find the brownian particle at time $t_w$
on site ${\bf x}$ is, in the long time limit:
\begin{equation}
{\cal P}({\bf x},t_w|{\bf 0},0) =
\frac{1}{(2 \pi)^{d}}
\left(\frac{2 \pi d \tau_0}{t_w} \right)^{d/2}
\exp{\left(- \left(\frac{d \tau_0}{2 t_w a^{2}}
{\bf x}^{2} \right) \right)}
,
\end{equation}
where the particle is at the origin at time $t=0$ (out of
equilibrium state). As noticed first in \cite{CKP},
the two times correlation functions of the position
are proportional to the waiting time:
\begin{equation}
\label{C(1)}
{\cal C}^{(1)}(t_w+t,t_w)=
\langle {\bf x}(t_w+t).{\bf x}(t_w) \rangle
=
\langle {\bf x}^{2}(t_w) \rangle
= \frac{t_w a^{2}}{d \tau_0}
\end{equation}
in the large $t_w$ limit.
In order to make contact with \cite{CKP}, one should
adopt the scaling $\tau_0 \sim a^{2}$ in the
thermodynamic/long time limits, which allows to recover the
Langevin dynamics results and thus the existence of ageing
in any dimension for the position--position
correlation function.

In order to make contact between the correlation
(\ref{C(1)}) and the general form of
the correlation (\ref{correlation}),
we should choose the overlap $q({\bf x}) =
{\bf x}^{2}$. With this choice of the overlap,
(\ref{correlation}) and (\ref{C(1)}) are both related
to the same physical quantity: the mean square displacement.
Using this specific form for the overlap,
one can calculate ${\cal C}_{\infty}(t)$,
integrate with respect to $t$ and show that
$\tau_w(\infty)=\infty$, in agreement with
the existence of ageing in this system for
the position correlations.

One could also be interested in the
following correlation:
\begin{equation}
\label{C(2)}
{\cal C}^{(2)}(t_w+t,t_w) =
\langle \delta_{{\bf x}(t_w+t),{\bf 0}}
\delta_{{\bf x}(t_w),{\bf 0}}
\rangle
= {\cal P}(0,t_w|0,0) {\cal P}(0,t_w+t|0,t_w)
.
\end{equation}
In the large $t$, $t_w$ limit, this correlation
is found to be
\begin{equation}
{\cal C}^{(2)}(t_w+t,t_w) =
\left( \frac{d \tau_0}{2 \pi} \right)^{d}
\frac{1}{(t t_w)^{d/2}}
.
\end{equation}
In the thernodynamic limit, $a \rightarrow 0$,
$\tau_0 \sim a^{2}$, so that ${\cal C}^{(2)}$
is vanishing in the thermodynamic limit,
signaling the absence of ageing for the correlation
(\ref{C(2)}).
The overlap associated to the correlation
(\ref{C(2)}) is the local one:
$q_{{\bf x}}=\delta_{{\bf x},{\bf 0}}$. We can thus
calculate $\tau_w(\infty)$ using the integration
(\ref{tauREM}) over the density of eigenvalues.
In order to calculate the behavior of $\tau_w(N)$,
we consider a finite system of volume
$V=L^{d}$ with a lattice spacing $a$.
The number $N$ of sites is
$N=(L/a)^{d}$. By thermodynamic limit,
we mean that $L$ is fixed, $a \rightarrow 0$
and $\tau_0 \sim a^{2}$.
Using (\ref{tauREM}), we have then
\begin{equation}
\label{tau(2)}
\tau_w^{(2)}(N) =
\left(\frac{a}{L} \right)^{d}
\int \frac{d^{d} {\bf k}}{(2 \pi/L)^{d}}
\frac{1}{\lambda_{\bf k}}
,
\end{equation} 
where the summation is restricted over
the wave vectors ${\bf k}$ such as
$1/L < |k_{\alpha}| < \pi/a$.
Clearly, the dominant contribution to
(\ref{tau(2)}) comes from the smallest
eigenvalues only. It is thus legitimate
to expand the eigenvalues up to the
quadratic order in ${\bf k}$:
\begin{equation}
\lambda_{\bf k} \simeq \frac{a^{2}}{\tau_0 d}
{\bf k}^{2}
.
\end{equation}
One then obtains
\begin{equation}
\tau_w^{(2)}(N) \sim
\frac{d \pi^{d/2}}{2^{d-1} \Gamma(d/2)}
\tau_0 a^{d-2} \int_{1/L}^{\pi/a}
k^{d-3} dk
.
\end{equation}
If $d=2$,
\begin{equation}
\tau_w^{(2)}(N) \sim
\pi \tau_0 \left( \ln{\left(\frac{\pi}{a}
\right)} - \ln{\left(\frac{1}{L} \right)} \right)
.
\end{equation}
Since $\tau_0 \sim a^{2}$, $\tau_w^{(2)}(N) \rightarrow 0$
in the thermodynamic limit in two dimensions.
Now if $d \ne 2$,
\begin{equation}
\tau_w^{(2)}(N) \sim \tau_0
\frac{d \pi^{d/2}}
{ (d-2) 2^{d-1} \Gamma(d/2)}
\left( \pi^{d-2} - \left(\frac{a}{L} \right)^{d-2}
\right)
.
\end{equation}
Since $\tau_0 \sim a^{2}$, this expression
goes to zero in the thermodynamic limit
whatever the dimension.

We have thus shown that, in the case of random
walks, and in the case of the correlations
(\ref{C(1)}) and (\ref{C(2)}), the
existence/absence of ageing in the
thermodynamic limit
is consistent with the general argument
of section \ref{general}.
\section{The BM model at zero temperature}
\label{BM-model}
\subsection{Diagonalization of the Glauber dynamics}
\label{zeroT}
We assume that the traps have been ordered such as
$E_1<E_2<...<E_N$.
Unlike the REM--like trap model, we do not make any asumption about
the trap distribution.
The finite temperature dynamics is
\begin{equation}
T_{i,j} = \frac{1}{N} \frac{1}{1 + \exp{(\beta(E_i-E_j))}}
\end{equation}
if $i \ne j$ and satisfies detailed balance.
The zero temperature form of the Glauber matrix
of this model is very simple since $T_{i,i}=-i+1$,
$T_{i,j}=0$ if $i>j$ and $T_{i,j}=1$ if $i<j$
at zero temperature.
The spectrum is $\{0,-1/N,...,(-N+1)/N\}$.
The eigenvector $|\psi_0 \rangle$
for the eigenvalue $\lambda=0$ is $|e_1\rangle$.
It is easy to show that the eigenvector $|\psi_{-k}\rangle$
associated to the eigenvalue $\lambda=-k/N$ is
\begin{equation}
\label{eig}
|\psi_{-k} \rangle = -\frac{1}{\sqrt{2}} |e_k \rangle
+ \frac{1}{\sqrt{2}} |e_{k+1} \rangle
.
\end{equation}
We can now express the pure states in terms of eigenvectors
\begin{eqnarray}
\label{dec1}
|e_1 \rangle &=& |\psi_0 \rangle\\
|e_k \rangle &=& |\psi_0 \rangle + \sqrt{2} \left(
|\psi_{-1} \rangle + ... + |\psi_{-k+1} \rangle \right)
.
\label{dec2}
\end{eqnarray}
This means that, at zero temperature, a pure state
$|e_k \rangle$ [$k$ is the $k$-ieth traps, with
$E_1<...<E_N$]
overlaps only with eigenstates
associated to eigenvalues $0,-1/N,...,(-k+1)/N$,
and this
overlap does not depend on the eigenstate. Compared
to the result of section \ref{eigI}, the structure
of the decomposition of a pure state into eigenstates
of the Glauber dynamics is very different from
the REM--like trap model.

If one starts from a pure state
$|e_k(0) \rangle$ at time $t=0$, the state of the
system at time $t$ is given by
\begin{equation}
|e_k(t) \rangle = (1-e^{-t/N}) |e_1 \rangle
+ \sum_{l=2}^{k-1} \left(e^{(-l+1)t/N} - e^{-lt/N} \right)
|e_l \rangle + e^{(-k+1)t/N} |e_k \rangle
.
\label{evol}
\end{equation}
We recover the fact that this model can
reach its ground state in a time of order $N$ at zero
temperature, whereas in the limit of zero temperatures,
and for the REM--like trap model,
the time spent in a trap $i$ is proportional to
$\exp{-\beta E_i}$, which goes to infinity if
$\beta \rightarrow 0$ and $N$ fixed.

\subsection{Two times autocorrelation functions}
The two times autocorrelation functions were calculated
in \cite{Mezard} by means of a Laplace transform formalism.
The result is
\begin{equation}
\Pi(t,t_w) \sim \frac{t_w}{t_w+t}
\end{equation}
if $1 \ll t,t_w$.
We rederive this result in the Appendix using a treatment
different from \cite{Mezard}, and based
only on the diagonalization of the Glauber matrix.
The BM dynamics thus exhibits ageing.
\subsection{Relation to the general argument}
Provided only self overlaps are non zero ($q_{i,j} =
\delta_{i,j}$), the correlation function introduced in
section \ref{general} is ${\cal C}_{\infty}(t) = \frac{1}{t}$,
provided $t \ll N$. The typical time $\tau_w^{BM}(\infty)$
below which ageing occurs is thus logarithmically divergent.
On the other hand, this can be also shown directly from the
spectrum:
\begin{equation}
\tau_w^{BM}(N) = \sum_{i=1}^{N-1} \frac{1}{n}
\end{equation}
which diverges logarithmically.
\section{One dimensional Ising chain}
As far as the finite temperature behavior of the
two times autocorrelation function is concerned, this
model exhibits interrupted ageing \cite{Bouchaud1}
since at any finite temperature the correlation
length is finite, of the order of $\xi = \exp{(2 J/T)}$,
ageing does not exist after a time independent of the
system size. It is thus interesting to apply to this case
the general argument of section \ref{general}.
We first begin with the definition of the Glauber
dynamics.
\subsection{Glauber dynamics of the one--dimensional
Ising chain}
We use the dynamics introduced in \cite{Glauber}. The master
equation of this single spin flip dynamics is
\begin{equation}
\frac{d}{dt} p(\{\sigma\},t) =
- \left( \sum_{i=1}^{N} w_i(\{\sigma\}) \right) p(\{\sigma\},t)
 + \sum_{i=1}^{N} w_i(\{\sigma_1,...,-\sigma_i,...,\sigma_N
\}) p(\{\sigma_1,...,- \sigma_i,..,\sigma_N\},t)
.
\end{equation}
As a consequence of detailed balance,
\begin{equation}
w_i(\{\sigma\}) =
\frac{1}{2} \left(1-\sigma_i \tanh{(\beta J
( \sigma_{i+1}+\sigma_{i-1})} \right)
.
\end{equation}
Notice that, unlike the case of the REM--like trap or BM models,
there is no need to multiply the matrix elements by $1/N$.
It is shown in \cite{Melin1} that in the infinite temperature
limit, the eigenvalues of the Glauber matrix are negative integers
between $-N$ and $0$, with degeneracies given by the
binomial coefficients. The largest relaxation time
in the infinite temperature limit is equal to unity
whatever the system size,
and there is thus no need to rescale the
coefficients of the Glauber matrix.

\subsection{Relation to the general argument}
In the case of a spin model, the overlap $q(\{\sigma\},
\{\sigma'\})$ is given by
\begin{equation}
q(\{\sigma\},\{\sigma;\}) = \frac{1}{N} \sum_{i=1}^{N}
\sigma_i \sigma_i'
.
\end{equation}
Since ageing is interrupted, we should observe a saturation
of the time $\tau_{w}^{Ising 1D}(N)$ as $N$ increases.
Moreover, $\tau_{w}^{Ising 1D}(N)$ should increase
as the temperature decreases. In practise, we work
with an open chain and we take advantage of the existence
of the global $Z_2$ symmetry (invariance of the
Glauber dynamics under the $\{\sigma\} \rightarrow
\{-\sigma\}$ transformation in a zero magnetic field),
and the reflection symmetry of the chain. The
Glauber matrix is diagonalized in each of these four
symmetry sectors. The results are plotted on figure
\ref{Fig5}. The difference $\tau_{w}^{Ising 1D}(N)
- \tau_{w}^{Ising 1D}(N-1)$ decreases as $N$ increases,
in agreement with the existence of interrupted ageing.

\section{SK model}
This model was proposed in 1975 as an "exactly solvable"
model of spin glasses \cite{SK}. For reviews on this model
and its developments, we refer the reader to the reviews
\cite{reviews}. The Hamiltonian of the SK model is
\begin{equation}
H = \sum_{\langle i,j \rangle} J_{i,j} \sigma_i \sigma_j
,
\end{equation}
where the sum is carried out over all the pairs of sites
(infinite dimensional model) and the quenched random interactions
are distributed according to  a gaussian distribution
\begin{equation}
P(J_{i,j}) = \left( \frac{N}{2 \pi} \right)^{1/2}
\exp{\left(- \frac{N J_{i,j}^{2}}{2} \right)}
.
\end{equation}
This model is a spin glass if $T<1$ and a paramagnet otherwise.
The two times correlation function were shown in \cite{MF1}
to exhibit ageing in the glassy phase. We are going to apply
our general argument of section \ref{general} to the case
of the REM--like trap model. To do so, we calculate numerically
$\tau_w^{SK}(N)$ for a large number of disorder realizations.
The result is plotted on figure \ref{Fig6}, as well as the
differences $\Delta_N = \tau_w(N+1)-\tau_w(N)$ for $N=3,4,5,6$ sites.
We observe that $\Delta_N$ increases at low temperatures
and decreases at high temperature. The crossing point on
the insert of figure \ref{Fig6} should be identified
to the glass temperature. This temperature is not equal
to unity since the system sizes are very small.
Notice that even though the sizes are quite small, a
large number of disorder configurations is required to
have a significant statistics (of the order of 500000
independent disorder configurations). The computer time
required was equivalent to 1000 hours of CPU time on
a single processor.
Our results
are thus consistent with the existence of ageing in the
low temperature phase, and no ageing in the high temperature
phase.
\section{Conclusion}
We have thus carried out a detailed analysis of ageing
from a Glauber dynamics point of view, with an
emphasis on finite size effects. We have shown how
to extract information from the Glauber matrix about ageing,
and how to analyze finite size effects. In some simple cases
with only self overlaps, the behavior of the time after
which there is no ageing only involves the knowledge
of the average density of eigenvalues $\rho_N(\lambda)$
when $\lambda \rightarrow 0^{-}$. This is the case for
the REM--like trap model,
in which case we have carried out a careful
analysis of the density of small eigenvalues. Our analysis
was confirmed by numerical diagonalizations of Glauber
matrices of small clusters. In this case, we were
able to related the typical behavior of the
density of eigenvalues to the spectral dimension
of the corresponding diffusion process.

\noindent \underline{Acknowledgements}:
R.M. acknowledges interesting discussions with
A. Barrat, J.P. Bouchaud, S. Franz 
and N. Wingreen. The authors acknowledge
B. Dou\c{c}ot for his interest to this work and
his encouragements. R.M. also acknowledges the
hospitality of NEC Research Institute at Princeton where
part of this work was done.

\section{Appendix}
We are going to calculate the two times autocorrelation
function $\Pi(t,t_w)$ using a method based only on
the exact diagonalization of the Glauber matrix at a
zero temperature. Starting from an infinite temperature
state at time $t=0$,
\begin{equation}
|\Psi(0) \rangle = \frac{1}{N} \sum_{i=1}^{N} |e_i \rangle
\label{initial}
,
\end{equation}
with $|e_i \rangle$ a pure state, the natural definition
of the two times autocorrelation function within this model is
\begin{equation}
\Pi(t,t_w) = \sum_{i=1}^{N} {\cal P}(i,t_w) \exp{\Delta_i t}
,
\end{equation}
where ${\cal P}(i,t_w)$ is the probability to find the system in
the trap number $i$ at time $t_w$ and $\Delta_i$ is the escape
rate from trap $i$, that is the diagonal element of the
Glauber matrix $\Delta_i=T_{i,i}$. This definition of
the waiting time dependent autocorrelation function
is similar to the one used in \cite{Bouchaud3}.
Using the expression (\ref{eig}) of the eigenvectors
and the decomposition (\ref{dec1}) and (\ref{dec2}) of
the pure states into eigenvectors, we can calculate
in a straightforward fashion the evolution of
the system, starting from the state (\ref{initial}):
\begin{equation}
|\Psi(t_w) \rangle = \left( 1 - (1-\frac{1}{N}) e^{-t_w/N} \right)
|e_1 \rangle + \frac{1}{N} \sum_{k=2}^{N}
e^{-k t_w/N} \left[
(N-k+1) e^{t_w/N}-(N-k) \right] |e_k \rangle
.
\end{equation}
The probabilities ${\cal P}(i,t_w)$ are easily recovered:
${\cal P}(i,t_w)= \langle e_i | \Psi(t_w) \rangle$.
After straightforward calculations, the dominant behavior
of $\Pi(t,t_w)$ is obtained, in the limit $1 \ll t,t_w \ll N$:
\begin{equation}
\Pi(t,t_w) \sim \frac{t_w}{t+t_w}
,
\end{equation}
which is equivalent to the result of \cite{Mezard}. Notice that
if $t_w$ is of the order of the number of traps, the
system has reached equilibrium: the dynamics is stationary
(the system is stuck in the lowest trap)
and there is no ageing.

\newpage
\renewcommand\textfraction{0}
\renewcommand
\floatpagefraction{0}
\noindent {\bf Figure captions}

\begin{figure}[h]
\caption{}
\label{Fig1}
Density of eigenvalues in the Arrhenius regime for the
REM--like trap model and fit to
the exponential form ($\beta=9$,$N=75$ traps). The insert
shows the same quantity, but with also the smallest eigenvalues.
For these small eigenvalues, the Arrhenius picture is not valid.
\end{figure}

\begin{figure}[h]
\caption{}
\label{Fig2}
Density of small eigenvalues of the models of section \ref{GM},
for $\beta=5$.
(a): infinite dimensional case (REM--like trap model);
(b): one dimensional case, with ``disordered'' traps;
(c): one dimensional case with ``ordered'' traps.
The lines in cases (b) and (c) are parallel,
and in good agreement with the Arrhenius law, whereas we observe
a significant deviation to the Arrhenius law in (b) case.
\end{figure}

\begin{figure}[h]
\caption{}
\label{Fig3}
Exponent $\alpha$ deduced from the fit (\ref{lawArr}) of
$P_B(\ln{|\lambda|})$.
The circles correspond to the infinite dimensional case
(REM--like trap model),
the squares to the one dimensional ``disordered'' case,
and the diamonds to the ``ordered'' one dimensional case.
The solid line corresponds the the Arrhenius
regime (in other words, to site--disordered
random walks above the upper critical dimension 2),
and the dashed line corresponds to the one dimensional
renormalized Arrhenius regime (in other words,
to site--disordered random walks in one dimension).
\end{figure}

\begin{figure}[h]
\caption{}
\label{Fig4}
Structure of the eigenstates of the REM--like trap model
for 20 traps. The label $i$ is an eigenvalue label, with eigenvalues
in growing order and $j$ labels the traps (the depth increases if $j$
increases).
(a) represents
$\langle \psi_i| e_j \rangle|$ for the $20$ pure states.
(b) represents the depth of the traps.
(c) represents the 20 eigenvalues between $-1$
and $0$ for 4 different temperatures (circles: $\beta=2$,
squares: $\beta=5$, diamonds: $\beta=10$, triangles: $\beta=20$).
\end{figure}

\begin{figure}[h]
\caption{}
\label{Fig5}
Variations of $\tau_w^{Ising 1D}(N)$ as a function of the inverse
temperature $\beta$ for $N=3,4,5,6,7,8,9$ sites. The insert
shows the differences $\tau_w^{Ising 1D}(N) -
\tau_w^{Ising 1D}(N-1)$ for $N=4,5,6,7,8,9$ sites
(curves a,b,c,d,e,f respectively).
\end{figure}

\begin{figure}[h]
\caption{}
\label{Fig6}
Variations of $\tau_w^{SK}(N)$ as a function of the inverse temperature
$\beta$ for a $N=3,4,5,6$ SK cluster. The insert shows
$\tau_w^{SK}(N) - \tau_w^{SK}(N-1)$ [a: $N=6$, b: $N=5$ c: $N=4$].
\end{figure}
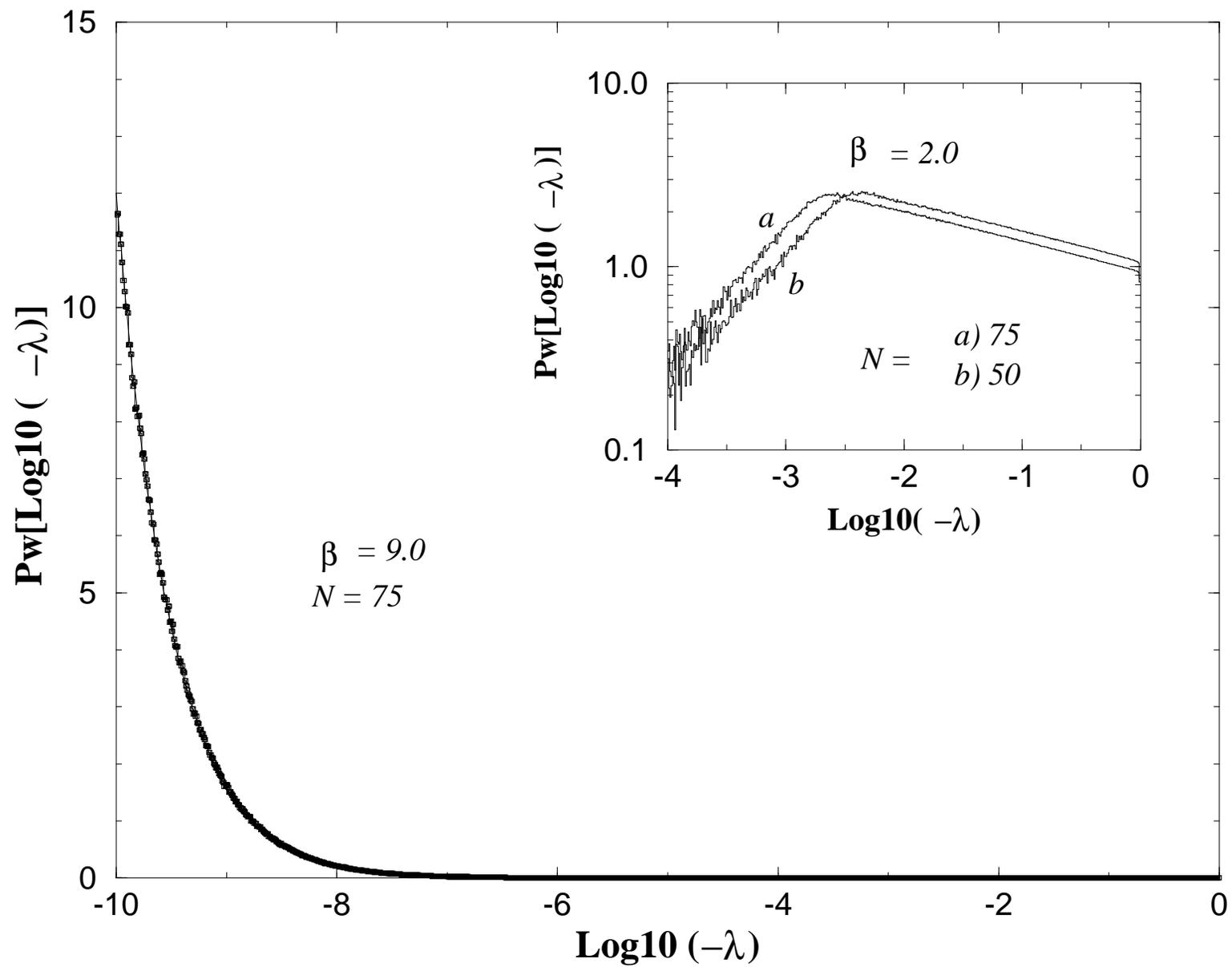
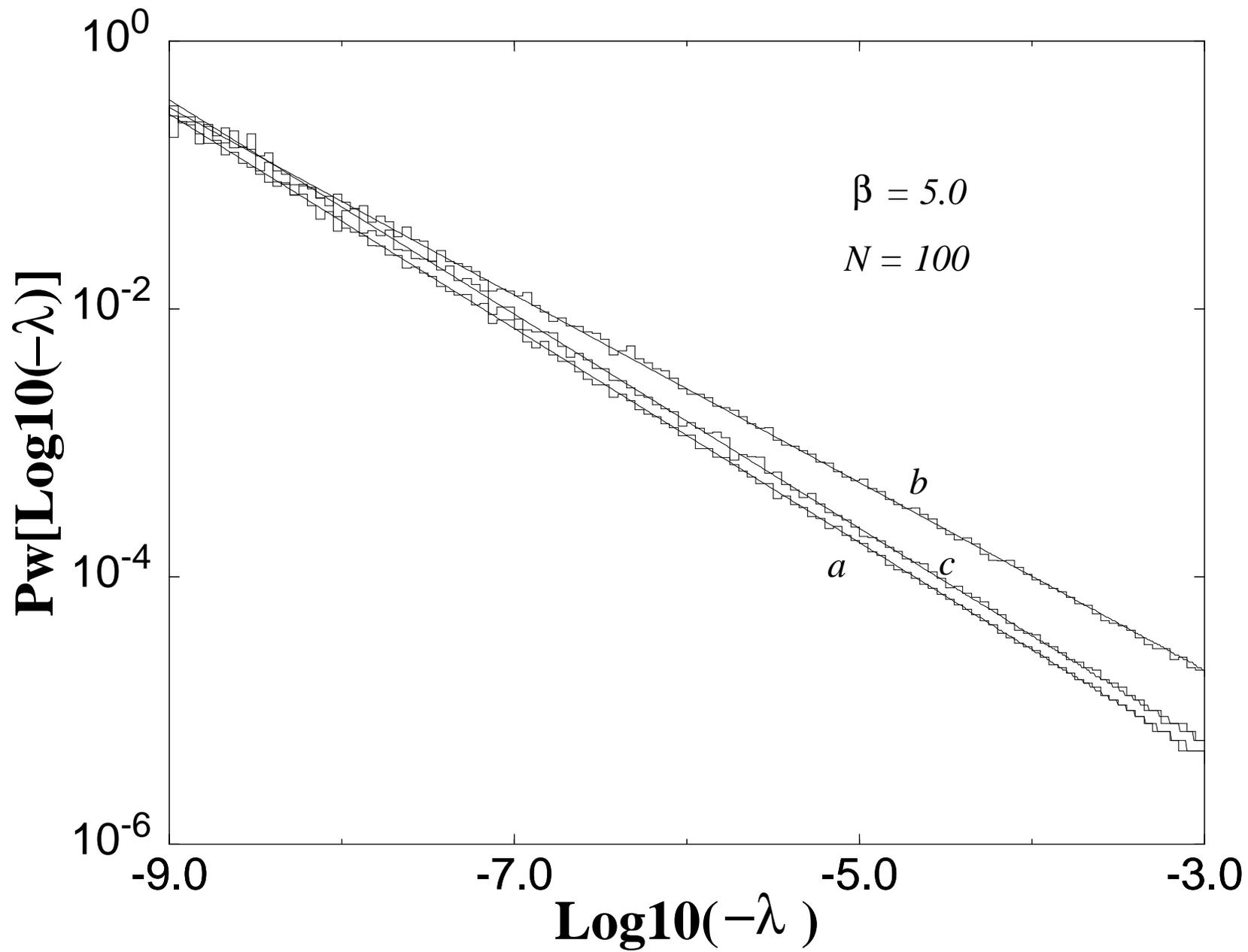
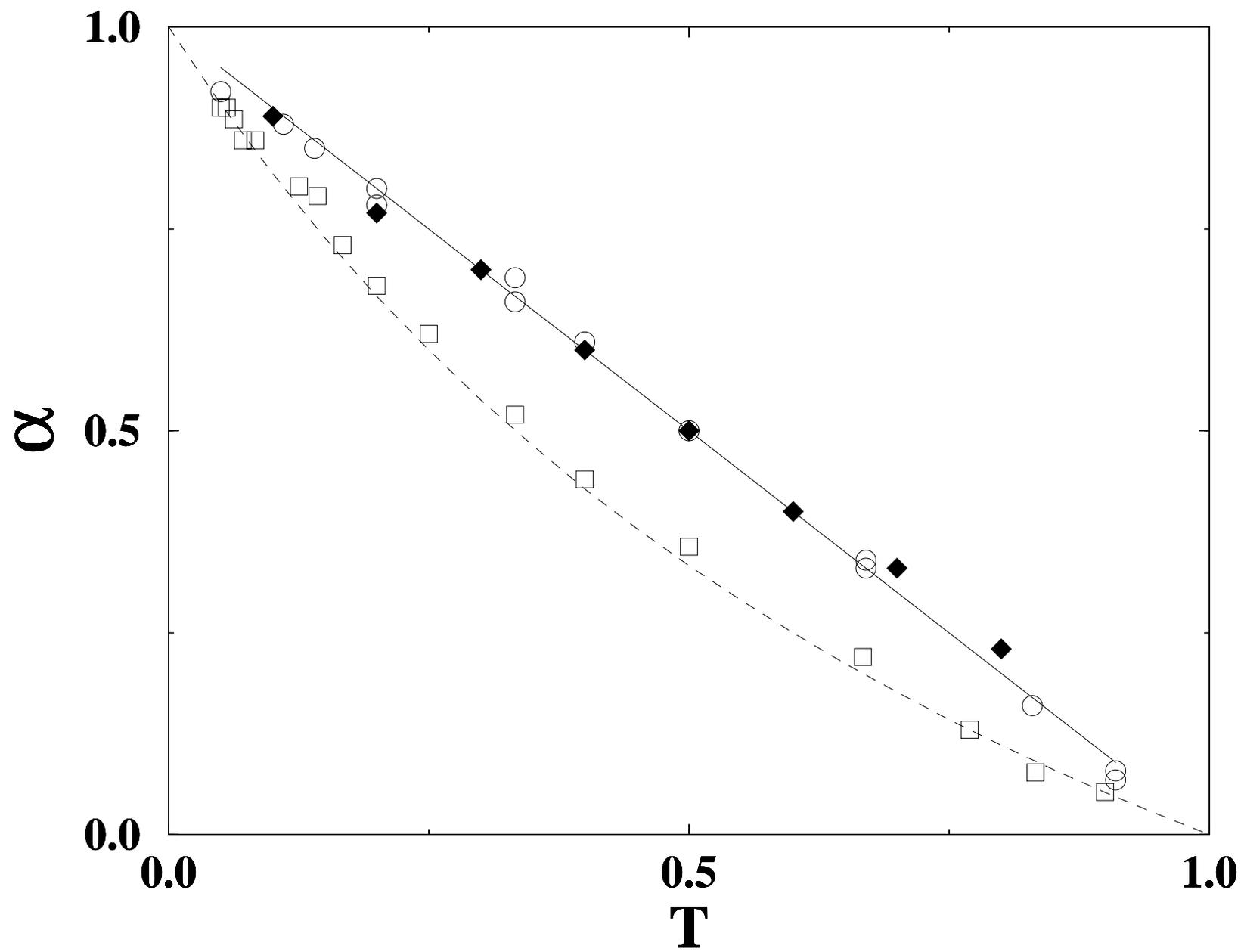
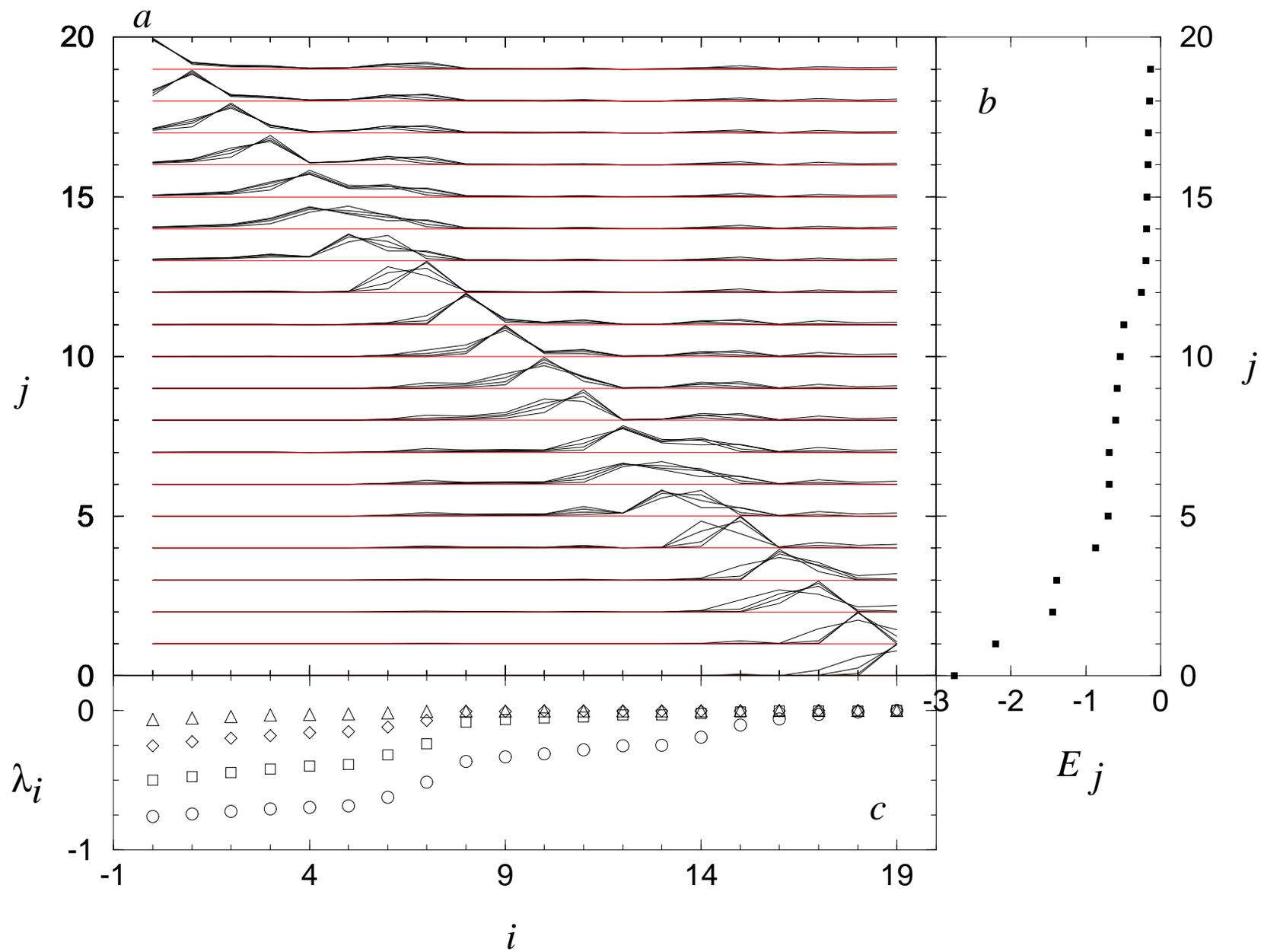
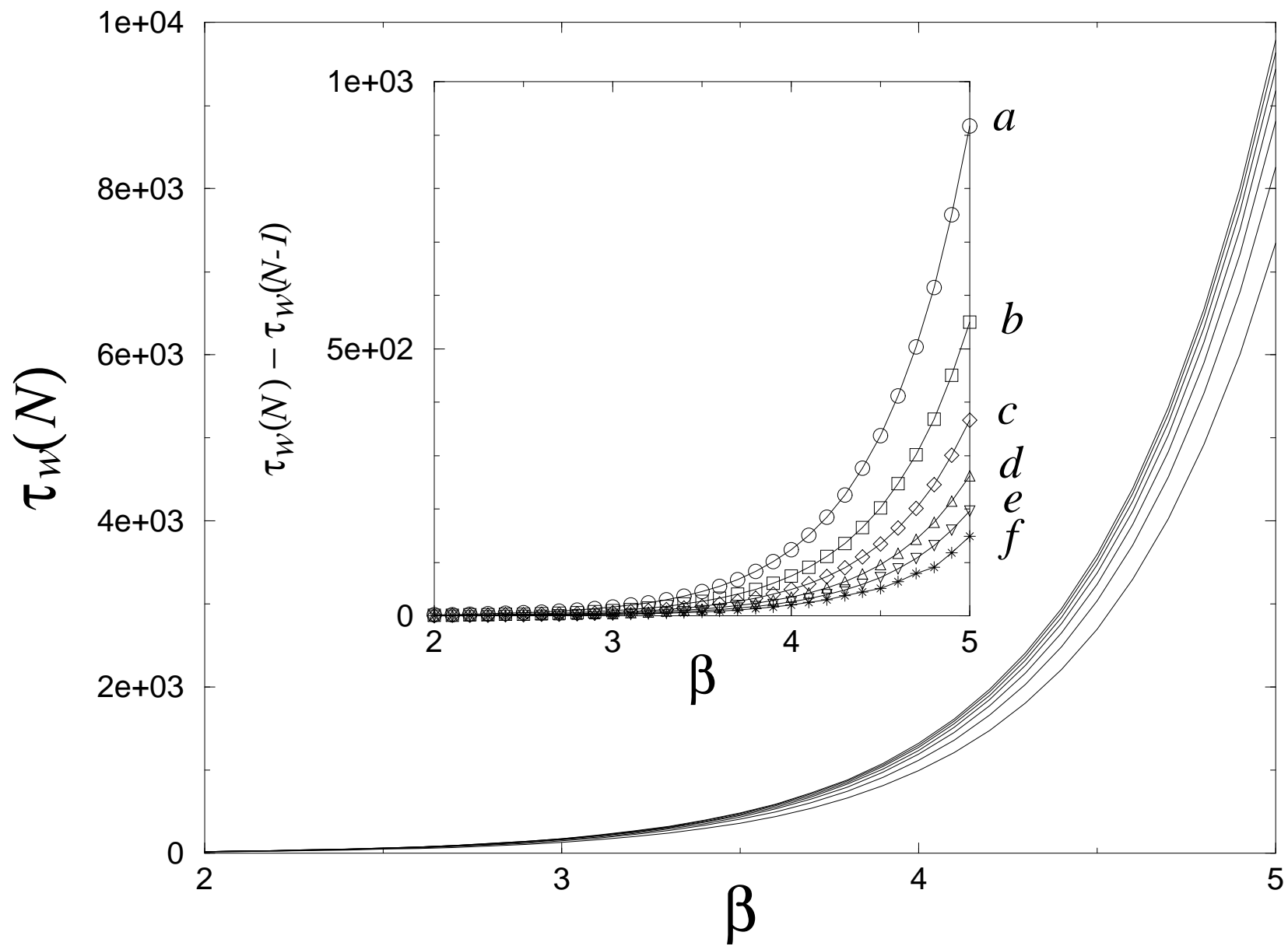
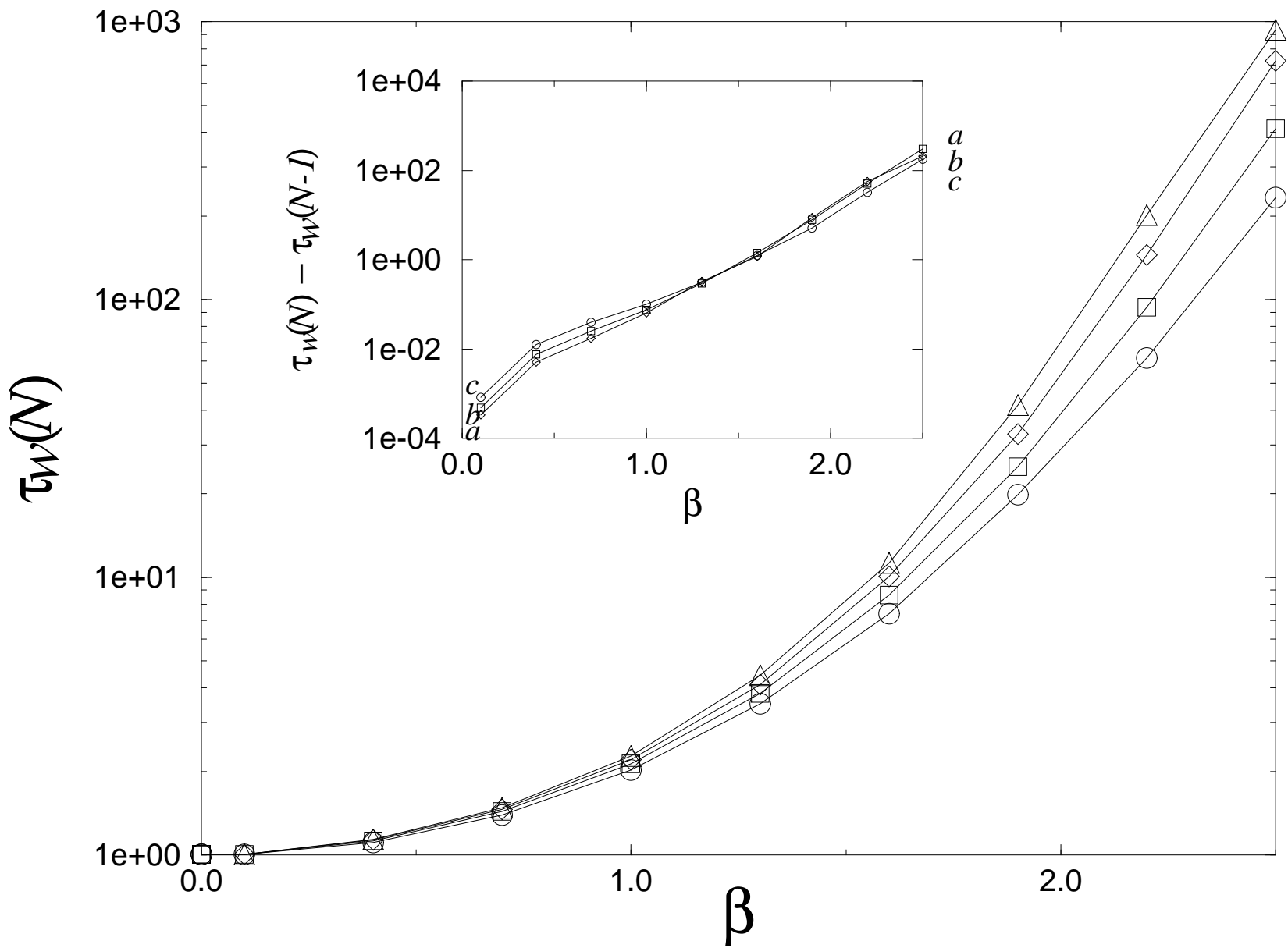
\end{document}